\documentclass[aps, prl, showpacs, reprint]{revtex4-1}

\usepackage{hyperref}
\usepackage{graphicx}
\usepackage{amsmath}
\usepackage{amsfonts}
\usepackage{amssymb}
\usepackage{bm}
\usepackage{color}

\newcommand{\mean}[1]{\mathinner{\langle \bm{#1} \rangle}}
\newcommand{\DeltaC}{\Delta_\mathrm{C}}
\newcommand{\DeltaA}{\Delta_\mathrm{A}}
\newcommand{\gammaperp}{\gamma_{\perp}}
\newcommand{\sgmj}{\sigma_{-}^\textrm{j} }
\newcommand{\sgpj}{\sigma_{+}^\textrm{j} }
\newcommand{\sgzj}{\sigma_{z}^\textrm{j} }
\newcommand{\driveE}{\mathcal{E}(t)}
\newcommand{\driveEnoT}{\mathcal{E}}
\newcommand{\Neff}{N_{\textrm{eff} }}
\newcommand{\etaT}{\eta}

\begin{document}

\title{Femtojoule-scale all-optical latching and modulation via cavity nonlinear optics}

\author{Yeong-Dae Kwon}
\author{Michael A.\ Armen}
\author{Hideo Mabuchi}
\email[Electronic address: ]{hmabuchi@stanford.edu}

\affiliation{Edward L.\ Ginzton Laboratory, Stanford University, Stanford, CA 94305}
\date{\today}

\begin{abstract}
We experimentally characterize Hopf bifurcation phenomena at femtojoule energy scales in a multi-atom cavity quantum electrodynamical (cavity QED) system, and demonstrate how such behaviors can be exploited in the design of all-optical memory and modulation devices. The data are analyzed using a semiclassical model that explicitly treats heterogeneous coupling of atoms to the cavity mode. Our results highlight the interest of cavity QED systems for ultra-low power photonic signal processing as well as for fundamental studies of mesoscopic nonlinear dynamics.
\end{abstract}


\maketitle

The diverse phenomena of cavity nonlinear optics~\cite{Mand05} provide a rich basis for fundamental studies of dissipative nonlinear dynamics~\cite{Lugiato:1984,Otsu00} and for the design of photonic signal processing devices~\cite{Gibb85}. Recent experiments exploiting resonant atomic nonlinearities~\cite{Kimble1991,Mabuchi:96,Chapman2004, Stamper-Kurn2007,Rempe2010,Kerckhoff:11} and/or nanophotonic cavities~\cite{Kimble2011,Painter2007,Vuckovic:2008,Notomi2012, Aspelmeyer2009, Kippenberg2010, Gaeta2013} have demonstrated that such phenomena can occur at very low (femtojoule--attojoule) energy scales in prototypical systems, reaching down to the quantum-physical few photon regime~\cite{Carm07,PhysRevA.73.063801} and raising intriguing prospects for corresponding ultra-low power photonic information technology~\cite{IBM_nano_photonics,Beausoleil2010,Miller:2010}. One of the less-explored dimensions of this developing scenario is the surprising complexity of dynamical behaviors, beyond simple thresholding and bistability, that can be achieved at low energy scales in cavities incorporating two-level atoms or comparable solid-state emitters.
In this Letter we present experimental data on sub-critical Hopf bifurcation phenomena in an atom-cavity system and demonstrate a novel device configuration that latches a logic level and provides rf modulation of the output signal at the same time.

The general paradigm of two-level emitters coupled to an isolated optical mode can be realized in a wide range of gas, liquid, and solid-state systems with various types of optical cavities; our experiment uses a 12~mm Fabry-Perot resonator in ultrahigh vacuum with an ensemble of laser-cooled $^{133}$Cs atoms placed between the mirrors. We utilize only the $|$F=m$_{F}$=4$\rangle$ to $|$F$'$=m$'_{F}$=5$\rangle$ transition of the D2 line at 852~nm where the atoms are well modeled as two-level systems. The following semiclassical equations of motion is used to model the system composed of a cavity and N weakly-coupled, noninteracting atoms~\cite{1071100}:
\begin{subequations} \label{eq:MBE}
\begin{align}
\dot{ \mean{ a } } &=
-(\kappa + i \DeltaC) \mean{a} + \sum_{j=1}^{N} g_j \mean{\sgmj} + \driveE \label{eq:MBE1}\\
\dot{\mean{\sgmj}} &=
-(\gammaperp + i \DeltaA) \mean{\sgmj}  +  g_j \mean{a} \mean{\sgzj} \label{eq:MBE2}\\
\dot{\mean{\sgzj}} &=
-2\gammaperp(\mean{\sgzj} + 1 ) -4 g_j \textrm{Re} \big\{ \mean{\sgpj}\mean{a} \big\} \label{eq:MBE3}
\end{align}
\end{subequations}
Here, $\bm{a}$ is the intracavity field annihilation operator, $\bm{\sigma_{\pm}}^j$ are the rasing/lowering operators for the $j$-th atom, and $\bm{\sgzj} = [\bm{\sgpj}, \bm{\sgmj}]$. The rates $g_j, \gammaperp, \kappa$ are the atom-cavity coupling strength for the $j$-th atom, the atomic polarization decay rate, and the cavity field decay rate, respectively. $\DeltaC$ and $\DeltaA$ are the cavity and the atom detunings relative to the drive (laser) frequency, defined as $\DeltaC = \omega_{\textrm{cav} } - \omega_{\textrm{las} } $ and  $\DeltaA = \omega_{\textrm{atom} } - \omega_{\textrm{las} }$, and $\driveE$ is the drive field amplitude.

The above Maxwell-Bloch-type equations are derived from an unconditional Jaynes-Cummings master equation in the same manner as in~\cite{PhysRevA.73.063801} (see also~\cite{PhysRevA.78.015801}), but with a slight generalization that incorporates distinct coupling strengths for the individual atoms. We note that the `factorization' approximations made in the derivation, such as $\mean{a \sgpj} \approx \mean{a} \mean{\sgpj}$, mean that entanglement between the atoms and the field is ignored. This is justified because the coupling strengths for individual atoms are small, while the combined coupling strength as an ensemble is still large enough to cause a strong nonlinearity.

By setting the time derivatives in Eqs.~(\ref{eq:MBE}) to zero one can solve for the steady state values of $\mean{a}$, $\mean{\sgmj}$, and $\mean{\sgzj}$. The independent variable $\driveEnoT$ is then expressed as a function of $\mean{a}$ to derive the input-output characteristic of the system, and the stability of any (hyperbolic) equilibrium point can be determined by local linearization. Alternatively, Eqs.~(\ref{eq:MBE}) can be numerically integrated to produce sample trajectories that reflect nonlinear dynamical phenomena such as limit cycles, chaos (\cite{PhysRevA.41.3975,PhysRevLett.95.093902}), etc. In this Letter we make use of both equilibrium and transient data to assess the accuracy of our theoretical modeling approach.

In our experiment, the atoms are first collected and cooled by a magneto-optical trap (MOT) formed on a gold-coated mirror surface~\cite{PhysRevLett.83.3398}. This surface MOT is located outside the cavity, 2 cm away from the cavity axis, where the cooling beams are not obstructed by the cavity mirrors. The atoms are then further cooled to sub-Doppler temperature ($<10~\mu$K) by polarization-gradient cooling, optically pumped to the m$_{F}$=4 state, and then loaded in the magnetic trap. The magnetic trap, formed by currents through copper wires ($100-200~\mu$m diameter) buried underneath the mirror surface, can be moved by shifting the currents~\cite{PhysRevLett.86.608}, and is used as a conveyer to transport atoms (up to $\sim10^6$) into the cavity mode. The number of atoms to be transported are adjusted by changing the initial MOT size. At the end of the transport process ($200-300$~ms), the atoms are released from the magnetic trap and a large uniform magnetic field is applied in the direction of the cavity axis, which ensures that the quantization axes for the atoms and the field are aligned in the presence of possible stray fields near the surface. The strong bias field also lifts the degeneracy among the Zeeman states (Zeeman shift: $2\pi \cdot 17~$MHz).  Once the atoms are in position, the input laser is injected from one side of the Fabry-Perot cavity and drives the TEM$_{00}$ mode. The beam that leaks out from the other side of the cavity serves as the output of the system. Due to the loss at the cavity mirrors and the existence of two possible directions for the leakage, we estimate $\etaT$, the probability of an intracavity photon making it to the output channel, to be 0.41. The output beam goes through an optical isolator and is then measured by a homodyne/heterodyne detection scheme~\cite{Kerckhoff:11}.
The maximum atom-cavity coupling constant, the atomic polarization decay rate, and the measured cavity field decay rate are $(\textrm{g}_0, \gamma_{\perp}, \kappa)/2\pi$ = (0.72, 2.6, 2.3) MHz, respectively.

When we apply the model to our experiment, we assume a large number of atoms distributed uniformly inside the cavity, so that the values of $g_j$ in Eqs.~(\ref{eq:MBE}) follow directly from the gaussian standing-wave profile of the resonant field mode. Under this assumption the atomic ensemble is characterized by a single parameter $N_{\textrm{eff} } \equiv \sum_{j = 1}^{N} (g_j / g_0)^2$, where $g_0$ is the maximum coupling constant at the anti-node center of the cavity. This simplification applies best for a cavity filled with a homogeneous dispersive/absorptive medium, a scenario for which extensive studies have already been performed using the Maxwell-Bloch Equations (MBEs) \cite{PhysRevA.43.6284,Lugiato:1984}. The method used here differs from the conventional analytical treatment of the MBEs in that we do not simplify Eqs.~(\ref{eq:MBE1})-(\ref{eq:MBE3}) using the approximation of collective atomic variables, which assumes, among other things, low absorption rate by the medium~\cite{Lugiato:1984}. We instead keep track of the evolution of $10^3$ individual atomic variables for better accuracy, at the cost of increased computational load.

\begin{figure}[t]
\begin{center}
\includegraphics[width=3.4in]{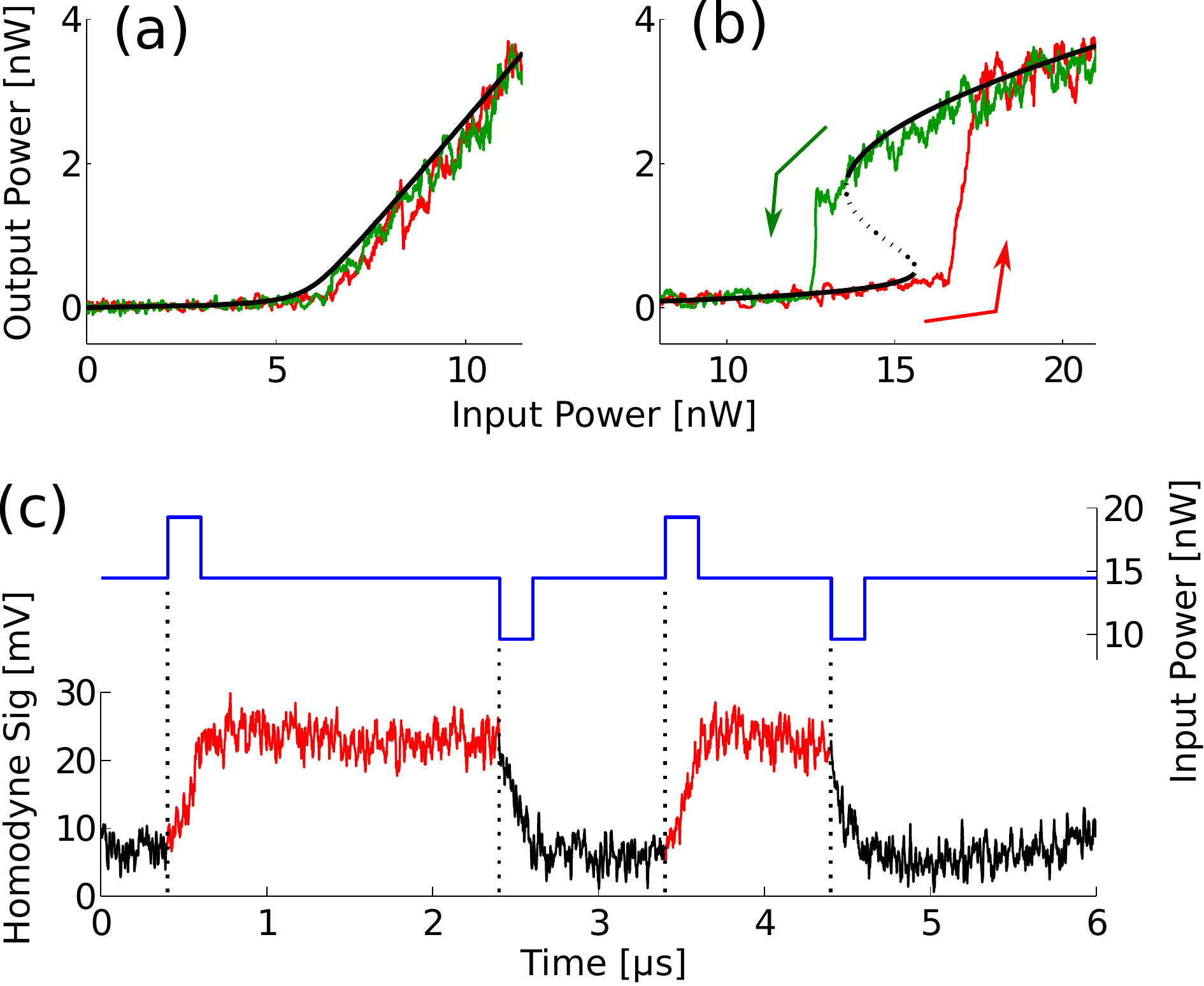}
\caption{%(color online).
(a)~Threshold behavior of the system with $\Neff = 130$, $\DeltaC = \DeltaA = 0$. The black line is the stable steady state solution derived from the model, and red (green) line is a typical data trace from the measurement when the input power is gradually increased (decreased).
(b)~Hysteretic behavior with $\Neff = 470$, $\DeltaC = \DeltaA= 2\pi \cdot 8~\textrm{MHz}$. Color scheme is the same as (a) except for the new black dotted line that indicates the unstable steady state solution.
(c)~Experimental demonstration of latching using bistable system parameters of (b). With the input power constant at 14.5nW the system maintains its current state. When a pulse is applied the system state is set high or low according to the pulse type (positive/negative).
}
\label{f:bistability_fig}
\end{center}
\end{figure}

The accuracy of our model in a regime of relatively simple dynamic behavior is illustrated in Figs.~\ref{f:bistability_fig}(a) and \ref{f:bistability_fig}(b), which display the measured equilibrium input-output response of our atom-cavity system for two different parameter sets $\{\Neff, \DeltaC, \DeltaA\}$ together with corresponding steady state solutions of Eqs.~(\ref{eq:MBE}). For the parameters of Fig.~\ref{f:bistability_fig}(a) ($\Neff=130$, $\DeltaC=\DeltaA=0$) a soft thresholding behavior is observed with a corner near 6~nW input power. With increased atom number and a detuned probe ($\Neff=470$, $\DeltaC=\DeltaA=2\pi\cdot 8~\textrm{MHz}$) as in Fig.~\ref{f:bistability_fig}(b) we observe classic bistability and hysteresis with coexisting stable equilibria from roughly 14 to 15.5~nW input power. For these measurements the input power was swept slowly (20~kHz) compared to the dynamical rates of the system ($\textrm{g}_j, \gamma_{\perp}, \kappa$) to probe the adiabatic response. Figures~\ref{f:bistability_fig}(a) and \ref{f:bistability_fig}(b) each show an individual (but precisely repeatable) data trace of $\sim50~\mu$s duration that was recorded continuously as the input power was swept up and down.

The theory curves in Figs.~\ref{f:bistability_fig}(a) and \ref{f:bistability_fig}(b) were computed using manually-adjusted values of $\Neff$ and $\driveE$ to improve agreement with our data, but the fitted values differ from the independently measured ones by no more than 10\%. $\Neff$ is recorded for each experimental run by measuring the spectral response of the atom-cavity system using a very low intensity probe~\cite{PhysRevLett.68.1132} right before and after the main measurement is made. Comparison of these two spectral curves also ensures that $\Neff$ does not vary significantly during the measurement period. This procedure and statement of calibration accuracy apply to all remaining data-theory comparison plots in this Letter as well. Some discrepancy between the model and the measurement is expected, mainly because the magnetic trap from which the atoms are released just before the measurement has a similar size to the cavity mode (waist $w_0 = 66~\mu$m), and therefore it is difficult to ensure that the atoms are truly uniformly distributed when the measurement is taken. Up to these $\sim 10\%$ inaccuracies however our model has proven to be a reliable predictor of the parameter values at which interesting nonlinear dynamical phenomena should be observed.

After confirming the parameters for simple bistable input-output behavior in our atom-cavity system, we performed an initial proof-of-principle experiment to demonstrate an all-optical set-reset (SR) latch, which could for example be used as an optical memory bit. Typical data is shown in Fig.~\ref{f:bistability_fig}(c). Here we envision a device configuration in which the set/reset control beam shares an optical input channel with the bias power beam (for example they could be combined by a beam splitter) so that they interfere constructively or destructively depending on the phase of the control beam. With this picture in mind, we demonstrated the set/reset control by superposing positive and negative pulses onto the laser beam that drives the atom-cavity system (upper blue trace of Fig.~\ref{f:bistability_fig}(c)). A positive pulse switches the response to the upper branch of the hysteresis curve, where it latches until a negative pulse resets it back to the lower branch. Our system required 15~nW optical bias power and the energy used for each control pulse was roughly $5~\textrm{nW}\times0.2~\mu\textrm{s} = 1~\textrm{fJ}$.

We note that the width of the control pulses used in our experiment corresponds to the inverse cavity decay rate $\kappa^{-1}$. Our switching time is thus longer than in nanoresonator-based devices, which generally have much larger $\kappa$ due to their small size. However, in terms of the switching energy, ours appears to be the minimum energy demonstrated in an all-optical bistable device~\cite{Notomi2012}. The switching energy potentially could be lowered even more as the energy scale in this experiment is still much larger than the optical shot noise (quantum fluctuation) limit~\cite{Kerckhoff:11}.

\begin{figure}[t]
\begin{center}
\includegraphics[width=3.4in]{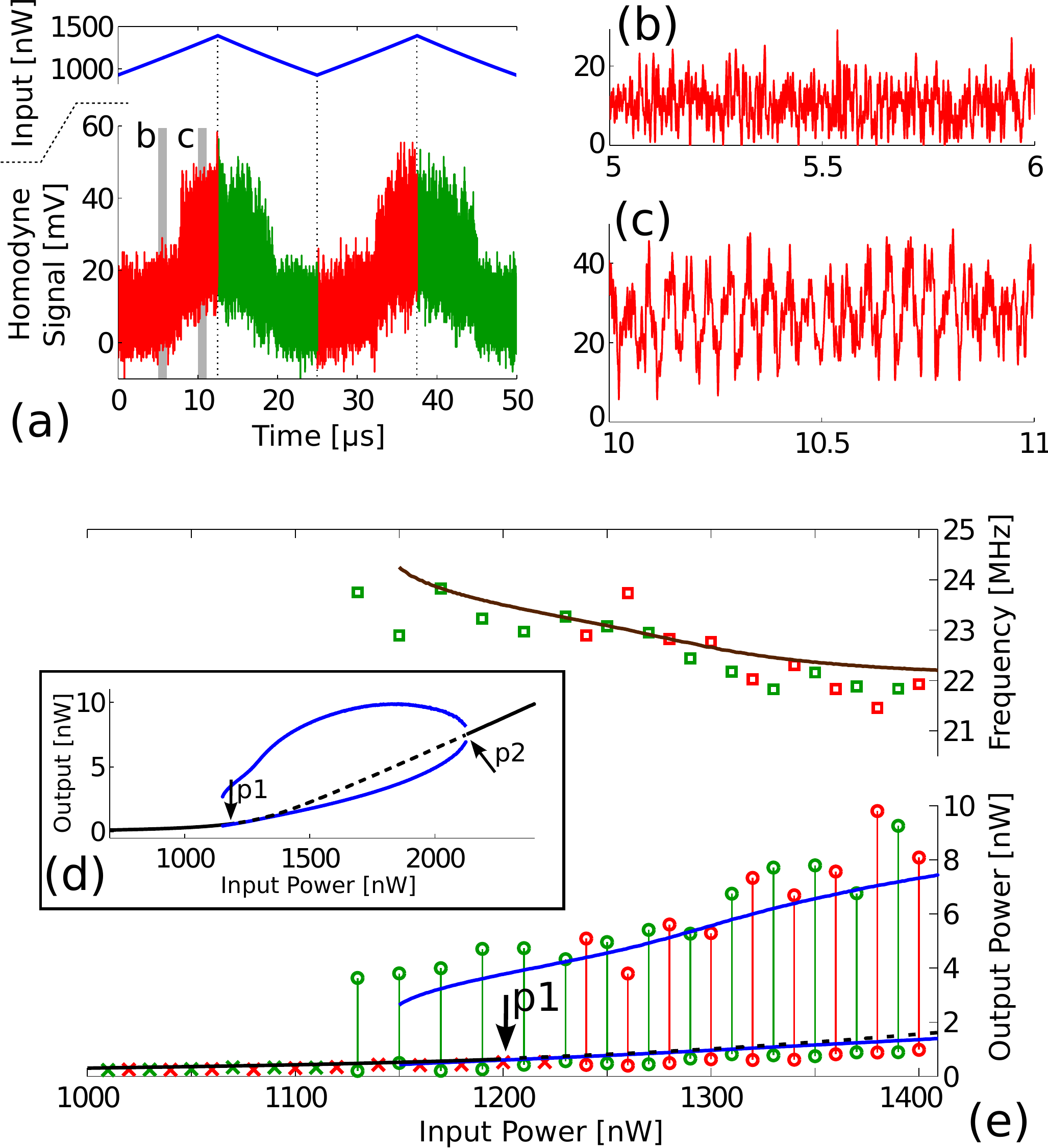}
\caption{%(color online).
Experimental parameters $\Neff = 2100$, $\DeltaC /  2\pi =  -20~\textrm{MHz}$, $\DeltaA / 2\pi  = 5~\textrm{MHz}$ are used.
(a)~Homodyne signal generated by the output beam as the input power is slowly swept up and down over the Hopf bifurcation point $\tt{p1}$. The gray bars labeled with `b' and `c' indicate the regions of which the zoomed-in views are displayed in (b) and (c).
(d)~Theoretical prediction of the input-output characteristics of the system. The solid and the dotted black lines represent the stable and the unstable steady state solutions, respectively. Points $\tt{p1}$ and $\tt{p2}$ indicate Hopf bifurcations discussed in the text. The upper and lower blue lines trace the maximum and the minimum output power of the limit cycles observed in simulations.
(e)~On the bottom, the predicted output power of the system (d) near the subcritical bifurcation point $\tt{p1}$ is compared with the measurement. The measured data is plotted in red (up-sweep) and green (down-sweep), using the following rules: When no oscillation is detected, the mean output power is recorded by a cross-mark. When the oscillation is detected, a sinusoid is fitted to the measured signal and its max/min optical power is denoted by a pair of connected circles.
On the top, we plot the frequencies of the both simulated and measured limit cycles. The simulation result is plotted in brown, while the measurements are denoted by red (up-sweep) and green (down-sweep) square marks.}
\label{f:hopf_bif}
\end{center}
\end{figure}

With an appropriate choice of external parameters $\Neff$, $\DeltaC$ and $\DeltaA$ guided by theory, we also probed the atom-cavity dynamics in an unstable regime and observed ``self-pulsing'' behavior ~\cite{Lugiato:1984, PhysRevA.30.1366} in which the global attractor is a limit cycle. In Fig.~\ref{f:hopf_bif}(a) we show the system making transitions between a stable equilibrium point and a limit cycle as we sweep the driving input power. When the input power is large, the power of the output beam oscillates significantly at a frequency in the range of $\sim 21-24$~MHz.

Our model predicts a range of input powers ($\sim 1150-1200$~nW) for which a stable equilibrium point and a stable limit cycle coexist. Starting from low input, as the system is driven past the sub-critical Hopf bifurcation point $\tt{p1}$ ($\sim$1200~nW, shown in Figs.~\ref{f:hopf_bif}(d) and \ref{f:hopf_bif}(e) the equilibrium point becomes unstable and the system jumps to the the limit cycle. With the system oscillating, the input power can be lowered back down and the limit cycle remains until another bifurcation occurs at a lower input power ($\sim$1150~nW) than $\tt{p1}$. As shown in Fig.~\ref{f:hopf_bif}(e), such hysteresis can be observed clearly in our experiment, and we find that the measured amplitude and frequency of the limit cycle closely match predictions. Moreover, combined measurements of the amplitude and phase quadratures of the output field confirm that the oscillation of the intracavity field follows a trajectory on the optical phase plane that is well predicted by theory (See Supplementary Fig.~S1).

Near the predicted super-critical Hopf bifurcation point $\tt{p2}$ ($\sim$~2100nW), on the other hand, the match between our measurements and theory was not as good. Simulation predicts that the amplitude of the limit cycle should gradually converge to zero as the system approaches $\tt{p2}$ from below, at which point a stable equilibrium point defines the steady-state. However, the experimental system continued to show large-amplitude oscillatory behavior at random intervals in time even when the input power was far above the predicted location of $\tt{p2}$. Simulations with added technical noise suggest that such behavior is quite reasonable, however, as the equilibrium points near $\tt{p2}$ are highly under-damped.

\begin{figure}[t]
\begin{center}
\includegraphics[width=3.4in]{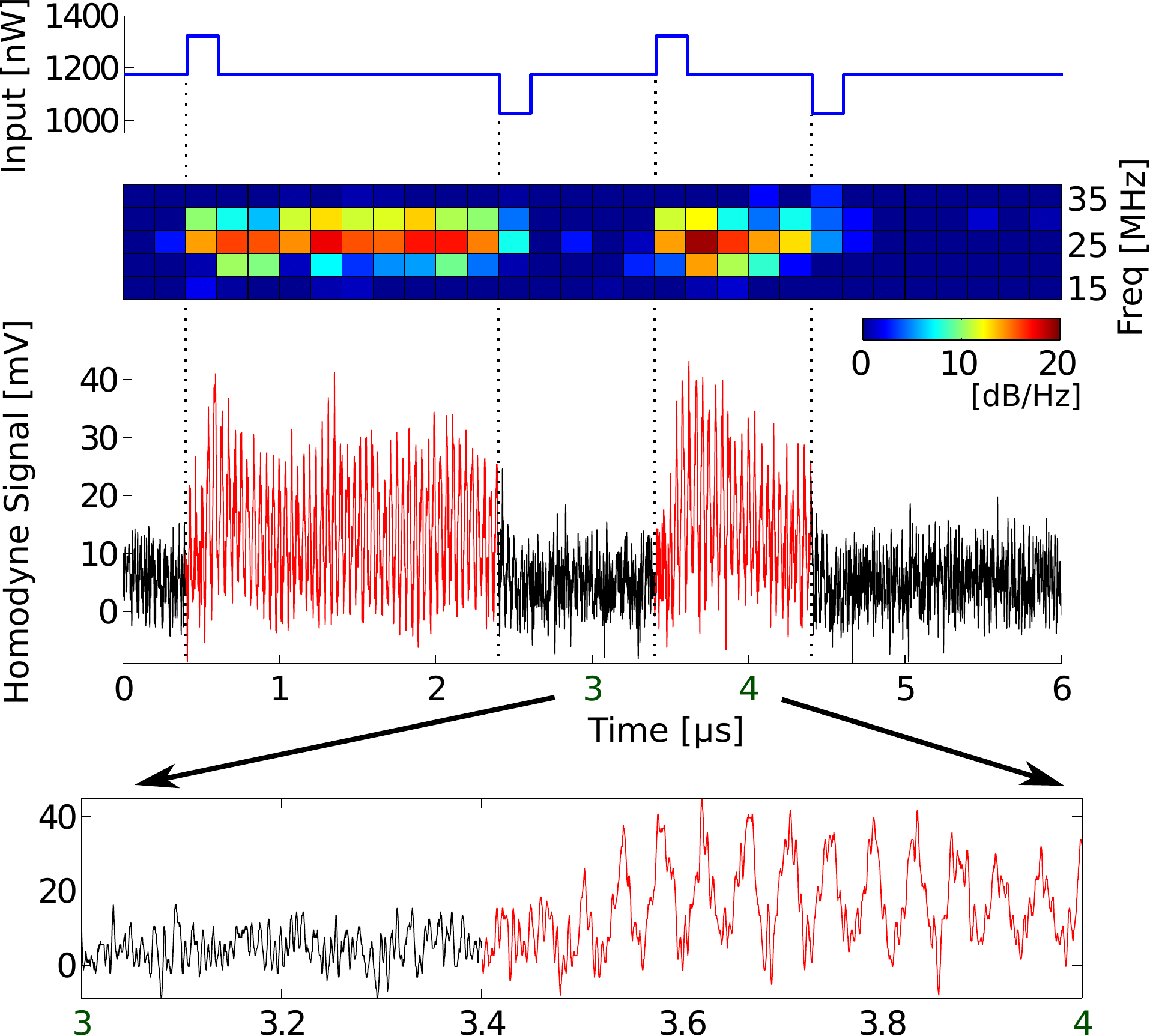}
\caption{%(color online).
The system shown in Fig.~\ref{f:hopf_bif} is used to demonstrate latching between a stable equilibrium point and a limit cycle.
The spectrogram in the middle is generated using $0.2~\mu$s segments of the homodyne data, which contain roughly 5 oscillation cycles. 0~dB power level was arbitrary chosen.
Note that more bandwidth is retained in the homodyne signal displayed here than in Fig.~\ref{f:bistability_fig}(c), in order to preserve the limit cycle oscillations. In the bottom frame we show a zoomed-in view of the homodyne signal from $3~\mu$s to $4~\mu$s.
}
\label{f:lc_latching}
\end{center}
\end{figure}

The limit cycle behavior of our system converts the constant input power into an rf-modulated output power. We are thus able to demonstrate an elementary physical mechanism that potentially could be adopted to make self-contained inline optical oscillators in a nanophotonic setting, for which the use of conventional optical-oscillator-type devices such as Q-switched lasers might not be practical. Our system furthermore is capable of transferring baseband power modulation to rf modulation of the output via the correlation between the input power and the frequency of the limit cycle (Fig.~\ref{f:hopf_bif}(e), upper plot). This suggests the possibility of an all-optical baseband-to-rf up-converter. And finally, hysteresis around the sub-critical Hopf bifurcation can be used to make a latch of rather unique characteristics, which switches between dc and modulated signal formats. Optical dc-rf latching devices may find use in optical communication systems where signal up-conversion is a common practice to avoid the high level of the background noise at low frequencies. Fig.~\ref{f:lc_latching} shows a demonstration of such latching performance. In this context we note that solid-state implementations of this type of cavity nonlinear optical device, for example using ensembles of quantum dots~\cite{Imamoglu:2007} or vacancy centers~\cite{Beausoleil2009}, could achieve much higher values of $(\textrm{g}_0, \gamma_{\perp}, \kappa)$ and therefore much higher limit cycle frequencies, potentially in the GHz range.

We find that the energy stored in the optical dc-rf latch in Fig.~\ref{f:lc_latching} is comparable to that of the elementary latch in Fig.~\ref{f:bistability_fig}, as in both cases the cavity contains about 500-1000 photons (100-200~attojoule) on average in the high energy state and much fewer when in the low energy state. However, the energy required for switching between the two states was more than an order of magnitude higher for the dc-rf latch ($\sim$30~fJ). This is mostly due to the mismatch between the input laser frequency and the resonances of the system.  In Supplementary Section Fig.~S2 we show a demonstration of sub-femtojoule dc-to-rf switching operation of the same Hopf bifurcation latch using a secondary near-resonant beam. We also note that the use of pulse shaping on the control pulse \cite{sandhu:231108} could potentially reduce the switching energy of both the elementary and the dc-rf latch further.

In conclusion, we have demonstrated that a cavity-based optical device that incorporates a resonant nonlinear response can exhibit complex dynamic behaviors such as hysteretic limit cycle formation at femtojoule energy scales. We have suggested novel signal processing functions based on such phenomena, and have demonstrated an all-optical configuration that simultaneously latches a logic level and performs rf modulation of the output power.
Intrinsic self-oscillatory behavior of the atom-cavity system, once a subject of active research using thermal gases and mW of laser power~\cite{first_limitcycle, PhysRevA.39.1235, PhysRevLett.60.412}, here has been probed with a laser-cooled atoms and $\mu$W power levels.  Analogous phenomena could be realized in the context of nanophotonics, where there is already an emerging interest in observing and functionalizing self-oscillations~\cite{PhysRevA.83.043816, PhysRevA.83.051802,GuT:2012}.

This work has been supported by the ARO (W911NF-08-1-0427) and DARPA (N66001-11-1-4106).

%%%%%%%%%%%%%%%%%%%%%%% References %%%%%%%%%%%%%%%%%%%%%%%%%

\bibliography{prl_bib.bib}

\end{document}